\documentclass[aps,preprint,nofootinbib]{revtex4-2}

\usepackage{graphicx}
\usepackage{amsmath}
\usepackage{amsfonts}
\usepackage{amssymb}
\usepackage{bm}
\usepackage{xcolor}
\usepackage{ulem}

\begin{document}

\title{Relation between diffusion equations and boundary conditions in bounded systems }

\author{F. Sattin}
\email{fabio.sattin@igi.cnr.it}
\affiliation{Consorzio RFX (CNR, ENEA, INFN, Università di Padova, 
Acciaierie Venete SpA), Corso Stati Uniti 4, 35127 Padova, Italy}

\author{Dominique Franck Escande}

\email{ dominique.escande@univ-amu.fr}
\affiliation{Aix-Marseille Université, CNRS, PIIM, UMR 7345, Marseille, France}

%\email{fabio.sattin@igi.cnr.it}
%\affiliation{
%Consorzio RFX (CNR, ENEA, INFN, Universit\`a di Padova, Acciaierie Venete SpA), 35127 Padova, Italy}

%\author{D. F. Escande}
%\email{dominique.escande@univ-amu.fr}
%\affiliation{
%Aix-Marseille Université, CNRS, PIIM, UMR 7345, Marseille, France}

\begin{abstract}
Differential equations need boundary conditions (BCs) for their solution. It is widely acknowledged that differential equations and BCs are representative of independent physical processes, and no correlations between them are required. Two recent papers [D. Hilhorst, et al, Nonlinear Analysis, {\bf 245}, 113561 (2024); J-W.Chung, et al, Jour. Math. Phys.  {\bf 65}, 071501 (2024)] argue instead that, in the specific case of diffusion equations (DEs) in bounded systems, BCs are uniquely constrained by the form of the transport coefficients. 
In this paper, we revisit how DEs emerge as fluid limits out of a picture of stochastic transport. We point out their limits of validity, and argue that in most physical systems, BCs and DEs are actually uncorrelated by virtue of the failure of the diffusive approximation near the system's boundaries. When, instead, the diffusive approximation holds everywhere, we show that the correct chain of reasoning goes in the direction opposite to that conjectured by Hilhorst and Chung: it is the choice of the BCs that determines the form of the DE in the surroundings of the boundary.  

\hfill

keywords: Fokker-Planck equation, Heat equation; Master Equation; Diffusion; Boundary conditions

\end{abstract}

\maketitle

\section{Introduction}
Diffusion equations (DEs) are ubiquitous in physics, employed for modeling the behaviour of stochastic or chaotic systems \cite{kotelenez08}. In this work we will refer specifically to the two expressions:
\begin{eqnarray}
\partial_t u   &=& \partial_x (D \, \partial_x u - V \, u) . \\
\partial_t u   &=& \partial_x (\partial_x (D \, u) - V' \, u) .
\end{eqnarray}
The two expressions above are obviously equivalent, related to each other through $V' = V + \partial_x D$. Historically, they were reached through different paths, though. Expression (1),  with $V = 0$, was first written by Fourier in 1822 to model phenomenologically the propagation of heat, and then by Fick, in 1855, to include also the transport of matter. Expression (2), with arbitrary $V'$, is the Fokker-Planck Equation (FPE), arising in the context of generic stochastic systems out of a microscopic picture of dynamics. 
Generally, $u$ represents any scalar quantity that evolves stochastically, $D$ quantifies the level of disorder, and $V$ accounts for any spatial asymmetry. 

Just like any differential equation, Eqns. (1,2) need to be supplemented by initial and boundary conditions (BCs) in order to be solved. In physics and engineering, one attributes different roles and meanings to the transport coefficients $D,V$, from the one side, and to the BCs to the other side: transport coefficients express the system's internal dynamics, while BCs represent the effect of external constraints. They are thus expressions of two independent physical mechanisms, hence no correlation is required to exist between the form of the DE and the kind of BCs that are imposed upon it. Arbitrary combinations of BCs and DEs are routinely employed in practical applications. \\
This view has been questioned recently by the two works \cite{bc1,bc2}. These papers examined a ``smooth'' version of bounded diffusive systems: The boundaries are not infinitely sharp; rather, a transition occurs between the inner region, characterized by  finite transport, and the outer region, where transport occurs with arbitrarily small coefficients $D,V$. 
The authors of \cite{bc1,bc2} argue that, {\it once that the analytical expression for $D$ and $V$ throughout all space is given}, only one ``natural'' kind of BC is consistent with the required boundedness of the motion. In particular, these are Neumann BCs with null derivatives for the Fick's equation (Eq. 1 with $V = 0$), and Dirichlet BCs with null density for the Fokker-Planck one without convection (Eq. 2 with $V' = 0$).  Thus, in the papers \cite{bc1,bc2}, DEs and BCs are not independent: the latter ones follow uniquely from the shape of the transport coefficients, and therefore from the DE.  These unexpected results call for a thorough re-examination of the subject of diffusive transport: it is the argument of the present paper.  \\
We note, first of all, that the range of validity itself of DE has to be scrutinized:  DEs are fluid models for physical processes that, ultimately, possess some sort of granularity. 
Section 2 is devoted, thus, to a re-examination of the conditions under which diffusion equations arise as appropriate descriptions of the physics. We recall that a microscopic picture of stochastic transport can be formulated as an ensemble of random walkers, whose dynamics is expressed in terms of jump probabilities $p$ from one site to another, and waiting times $\tau$ between successive jumps. The corresponding equation that quantifies the transition rates from different sites is often referred to in literature as ``Master Equation'' (ME). It is commonly argued that DEs arise as long-wavelength limits of the ME: this is the essence of the popular Kramers-Moyal (K-M) expansion.  On the other hand, the K-M expansion has been recognized since long to be based upon heuristic, not rigorous arguments. Therefore, we devote section 3 to a critical review of the arguments that have been advanced along the years to place the K-M result upon firmer grounds, fundamentally based upon the Central Limit Theorem. In section 4, armed with these results, we return to our main theme: the connection between DEs and BCs.  Our conclusions are: (i) in most systems, boundaries are not smooth, rather the interface between the interior and the exterior of the system is narrower than the average random walker' step. This introduces a discontinuity: the DE formalism cannot be extended from the core to the boundary, therefore making effectively independent DEs and BCs. (ii) In the presence of smooth boundaries, BCs  and DEs may instead be coupled, but in this case it is the hypothesis of the independence of the DE to fail; rather, we will show that the DE--near the boundaries--must adapt precisely to the BC, so as to preserve consistency.  

\section{DEs as fluid limits of stochastic dynamics}
We write the local rate of change of the scalar quantity $u$, $\partial_t u(x,t)$,  in terms of processes that move $u$ across different points according to some probability \cite{risken}:
\begin{equation}
\partial_t u(x,t) = \int { p(x,x') \over \tau(x')} u(x',t) dx' - {u(x,t) \over \tau(x)} .
\label{eq:me}
\end{equation}
In this expression, a version of a Master Equation (ME), $p(x,x') dx'$ is the probability for the scalar $u(x',t)$ to jump to $x$.  Consistently with most of the literature, we will suppose that $p$ depends from $x,x'$ through the combination: $p = p(\Delta = x-x'; x')$, i.e, it depends on the starting point and from the relative distance to the arrival point. The integral in Eq. (\ref{eq:me})-- implicitly understood as spanning all space--becomes thus a convolution. \\
Within this simplified picture of the dynamics, the time needed for $u$ to move from $x'$ to $x$ is regarded as negligible but, between any two consecutive jumps, there is a waiting time $\tau(x')$--which we suppose to depend upon the starting location $x'$ alone, still adhering to most of the literature. Additional effects may be added to Eq. (\ref{eq:me}). Among them: temporal memory effects, finite traveling times, sources/sinks, etc ..., but here we will stick to this barebone picture for stochastic motion. Throughout this work, we will consider just one-dimensional dynamics, but it is clear that Eq. (\ref{eq:me}) may generalized to higher dimensions, and also be regarded as an effective one-dimensional equation, where the dynamics along the other directions has been averaged out. See \cite{langevin,tabar} for further details.  \\
The integro-differential equation (\ref{eq:me}) can be turned into a fully differential one under some conditions.  The popular Kramers-Moyal (K-M) method \cite{risken,tabar} expands Eq. (\ref{eq:me}) in series of the small parameter $\Delta$, and truncates the expansion after the first two terms. Neither the series expansion nor the truncation to a finite number of terms are trivially justifiable, and a lot of attention has been devoted to them. We will return in the next section upon this issue in order to provide a firmer ground to the K-M result. \\ 
We arrive, eventually, to the FPE (2), where
\begin{equation}
D = \int d\Delta {\Delta^2 \over 2 } {p(\Delta,x) \over \tau(x)} , \quad V' = \int d\Delta \, \Delta \, {p(\Delta,x) \over \tau(x)}  .
\label{eq:dv}
\end{equation}
Important particular cases correspond to  $V' = 0$ and $V'= dD/dx $. 
In terms of stochastic processes, the condition $V' = 0$ is realized by jumps equally probable in both directions: $p(-\Delta,x) = p(\Delta,x)$. The condition $V' = dD/dx$, which--as we have remarked earlier--leads to Fick's form of diffusion,  is associated with a more complicated relation:
\begin{equation}
{p(\Delta, x) \over \tau(x)} = {p(-\Delta, x + \Delta) \over \tau(x+\Delta)} .
\label{eq:neumann}
\end{equation}
This expression is odd-looking at first sight, but actually is just an expression of a detailed balance, ensures that the rate of jumps from $x$ to $x'$ and {\it vice versa} are identical \cite{pla08}.  We recall that it was originally argued by Landau that the constraint $ V'  = dD/dx$ must hold for general 1 degree-of-freedom Hamiltonian dynamics, whereas it vanishes for higher dimensional dynamics \cite{prl07,ppcf08}. \\

\section{From the Master Equation to the Diffusion Equation, and back}
The mutual relation between the ME and the DE is actually a nontrivial topic, that has been--and to some degree still is--debated in the literature.  \\ 
In the previous section, we have justified the passage from the ME to the DE employing the popular Kramers-Moyal approach.  It amounts to expanding the ME in powers of the jump length $\Delta$, and truncate it after the second-order term. The K-M expansion is implicitly based upon the ansatz that a separation exists between  $\Delta$  and the slower variation scale of the density $u$. The lack of a rigorous control parameter which may provide a justification for the neglect of the higher-order terms has always been a source of concern. The Pawula theorem \cite{tabar,pawula} states that if the number of nonvanishing terms of the series is finite, it must be either one or two, in order to avoid unphysical features in the solution, such as negative density. This is, however, just a justification {\it a posteriori} of the K-M result. \\
In order to overcome this lack of rigour, van Kampen \cite{vankampen1,vankampen2} suggested his system-size expansion in terms of a small parameter $\Omega^{-1}$, where $\Omega$, often (but not necessarily) identified with the size of the system, is a manifestly large parameter, exceeding all other scales of the system. van Kampen  distinguishes a macroscopic deterministic  contribution to the system's trajectory, scaling like $\Omega$, and a fluctuating part, scaling like $\Omega^{1/2}$. 
van Kampen's approach is more rigorous than K-M's, but still contains a dose of arbitrariness, since the relative weight between macroscopic deterministic and fluctuating terms is set {\it ad-hoc}. Basically, his ansatz is based upon the law of larger numbers. Attempts of improving upon van Kampen's approach appear still nowadays (see, e.g., \cite{cianci17,peralta18}). However, a convincing analysis was carried out by Ryskin \cite{ryskin97},  where he recalls how a Markovian process, over times longer than the decorrelation time, fulfills the conditions for the validity of the Central Limit Theorem, and entails therefore the emergence of Gaussian processes. Ryskin's work explains why, even if the higher-order terms in the K-M expansion are not vanishing, they nonetheless may be disregarded: the reason is that they do not enter the long-time evolution equation \cite{nota}. In conclusion, the DE is a faithful picture of the dynamics only over times longer than the decorrelation one. Over these time scales, several random steps accumulate. The DE is therefore accurate only over distances larger than the jump length not because of the original Kramers-Moyal scale separation hypothesis, but because over short spatial scales the conditions for the validity of the Central Limit Theorem cannot be fulfilled.

\section{On the relation between DEs and BCs}
The previous section brings an important consequence for this work: transport around sharp structures is not faithfully described by the DE. In the case of neutral fluids bounded by material walls, the interface region is provided by the outermost atomic layer of the solid surface, and is therefore on the atomic scale, much smaller that any turbulence scale. In the case of bounded plasmas, there is some ambiguity because of the existence, on the one side, of several physically relevant lengths (Debye sheath, Scrape-Off-Layer) over which transport is modified and, on the other side, of several scales for the turbulence. 
The ME, on the other hand, does not suffer from the same problem: one is allowed to pick up arbitrary $p, \tau$, provided that they fulfill the coarse-scale constraints dictated by the experiment (This is easily understandable within the framework of the Bayesian approach to statistics), hence issues of scale are of no concern here: the reader is referred to \cite{pla08} for the explicit treatment of a case where ME and DE may yield different results. \\
As far as boundaries are sharp, therefore, the arguments of Hilhorst, and Chung cannot apply, since there is an interface region where the DE does not hold. This provides an effective discontinuity between the core and the boundary that invalidates their argument.  \\
Now, we show that the alleged inconsistencies invoked by Hilhorst and Chung do not arise even in the presence of a smooth transition between core and boundaries. The BCs encode physical information that is captured in the transport coefficients $p, \tau$ of the ME, and then transmitted to the diffusion coefficients $D,V$ of the DE in a way that must preserve consistency. Let us see how this occurs.  For example, let us start with a system whose dynamics, far from the boundaries, is left-right symmetrical: $p(-\Delta,x) = p(\Delta,x)$: $V' = 0$, and the pure FPE (2) holds.
When one approaches the boundaries, i.e., when the distance from the boundary is of the order of the average jump length, different scenarios may be envisaged, depending upon the boundary conditions holding. Let us start by considering absorbing boundaries: The walker may cross the boundary but, once there, it does not return into the system any longer. This scenario can be modeled with $\tau(x) $ going to infinity when $x$ is outside the boundary, while being finite inside the boundaries. The left-right symmetry properties of $p$ need not be affected, hence  $V' = 0$ still holds everywhere.
Let's consider now the case of reflecting boundaries. Each walker hitting the wall must return back inside the system. This is accomplished provided that the rates of outbound and inbound hops equilbrate: it corresponds to the condition (\ref{eq:neumann}) which leads to the Fick's form for the diffusion equation. 
In summary, while the DE, in the interior of the system, may take an arbitrary expression, near the boundaries it must adapt itself to a form that is consistent with the BCs. The width of this transition layer is likely of the order of the average jump length, hence microscopic and invisible to the experimentalist.   
For modeling purposes the ``core'' expression of the DE is the only one that can be compared with experiments, the ``edge'' expression of the DE escapes observations. In practical terms, the edge region remains a zone of discontinuity, as long as the diffusive approximation is invoked, just like in the scenario with sharp interfaces.   \\   
To conclude, we mention that--in principle--BCs may incorporate further constraints that cannot be captured into transport properties; for instance: external reservoirs. It is clear that in this case, too, the alleged dependence of the BCs from the DE is untenable.  \cite{nota2}.     

\section{Conclusions}
Basically, our paper is aimed to explain how the conclusions drawn in the works \cite{bc1,bc2} are not physically applicable, since they postulate that one may employ the Diffusion Equation regardless of any other constraint. We have shown that it is not the case. We have recalled that the DE is a valid picture of the physics of transport only over sufficiently large length scales. When the boundary-system interface width is smaller than these scales, it acts as an effective discontinuity region, making independent the physics determining the DE, valid in the interior, and that determining the BCs. When the transition region is large enough to make the diffusive approximation valid throughout the whole system, on the other hand, the DE and the BCs both arise consistently out of the same model of transport quantified by the ME. This entails that, in the transition region, the DE cannot be given independently from, rather must adapt precisely to the form consistent with, the BCs. In both cases, we re-establish the common notion that BCs do not depend from the DE. \\
A spin-off of our study--that we regard as valuable in a journal devoted to the foundations of the physical theories--is that we were forced to revisit the historical development that has led to the Diffusion Equation, highlighting some lesser known results (Ryskin’ analysis of the K-M result as arising from the Central Limit Theorem), as well as correcting some misinterpretations (Ryskin’ claim of a precedence of the DE equation over the ME).

\end{document}